%% file: skeleton.tex
\documentclass[a4paper,11pt]{article}
\usepackage{pos}

\usepackage{natbib}
\title{Study of multiple ring ELVES with the Mini-EUSO telescope on-board the International Space Station}
 \ShortTitle{ELVES with Mini-EUSO}

\author*[a,b]{Giulia Romoli}

\affiliation[a]{University of Rome Tor Vergata, Via della Ricerca Scientifica 1, Rome, Italy}

\affiliation[b]{Istituto Nazionale di Fisica Nucleare - Sezione di Roma Tor Vergata, Rome, Italy}

\onbehalf{for the JEM-EUSO Collaboration\\[-1mm]{\normalsize \normalfont (a complete list of authors can be found at the end of the proceedings)}}

\emailAdd{giulia.romoli@roma2.infn.it}

\abstract{Mini-EUSO (Multiwavelength Imaging New Instrument for the Extreme Universe Space Observatory) is a telescope observing the Earth in the ultraviolet band (290-430 nm) from the Russian Zvezda module of the International Space Station since 2019. The telescope is capable of observing UV emissions of cosmic, atmospheric, and terrestrial origin on different time scales. Among the atmospheric phenomena that can be studied, ELVES (Emission of Light and Very low-frequency perturbations due to Electromagnetic pulse Sources) have been photographed by Mini-EUSO with a time resolution of 2.5 µs. ELVES are rapidly expanding rings of optical and ultraviolet emissions, 75-95 km in height, resulting from the de-excitation of molecular nitrogen and oxygen in the lower ionosphere following a lightning-associated ElectroMagnetic wave Pulse (EMP). A detailed study of their characteristics, such as radius, speed, and energy, is required for the understanding of these phenomena. In this work, results from the observation of about 30 ELVES with Mini-EUSO will be presented. Using dedicated algorithms, their electro-optical dynamics and morphological characteristics have been thoroughly investigated.}

\ConferenceLogo{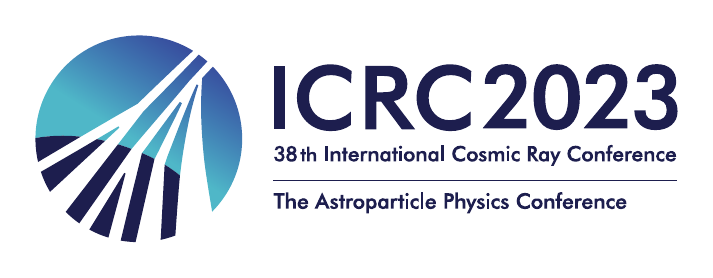}

\FullConference{%
38th International Cosmic Ray Conference (ICRC2023)\\
  26 July - 3 August 2023\\
  Nagoya, Japan}

\begin{document}
\maketitle

\section{Introduction} 

Mini-EUSO (Multiwavelength Imaging New Instrument for the Extreme Universe Space Observatory) \cite{Bacholle2021} is a telescope operating on board the International Space Station (ISS) since 2019, that observes the Earth from a nadir perspective in the ultraviolet, visible and near-infrared ranges (predominantly between 290 and 430 nm). The detector is part of the international JEM-EUSO (Joint Experiment Missions for Extreme Universe Space Observatory) collaboration, which aims to investigate Ultra High Energy Comic Rays (UHECRs) from space, 
by observing the fluorescence and the Cherenkov light from related Extensive Air Showers (EAS) in the Earth’s atmosphere \cite{Casolino2017}. 

The primary objectives of Mini-EUSO encompass a wide range of scientific investigations. The telescope observes both natural and artificial terrestrial ultraviolet emissions, including clouds, marine bioluminescence, and nocturnal city lights. In this context, it provided the first UV night-time maps of the Earth \cite{Casolino2023}. Additionally, the telescope is involved in the study of atmospheric phenomena, such as Transient Luminous Events, as well as in the detection of meteors, meteorites, and space debris. Finally, it aims to contribute to the search for nuclearites and Strange Quark Matter. 

The telescope's spatial resolution of $\simeq4.7\;\text{km}$ at the ionospheric level, combined with a temporal resolution of $2.5\;\mu\text{s}$, has proven to be well suited for the detection of ELVES (Emission of Light and Very low-frequency perturbations due to Electromagnetic pulse Sources)\cite{Marcelli2021}. ELVES are transient flashes of light expanding radially in the lower ionosphere (75 - 105 km altitude), as rings of increasing radius \cite{pasko2012}. They originate from electromagnetic pulses radiated by lightning return stroke currents, which expand upwards from the lightning cloud and cause electron heating and subsequent de-excitation of molecular species in the lower ionosphere, such as nitrogen and oxygen. Almost all ELVES have a distinct hole in the center due to the shape of the dipole radiation, and expand to radii greater than 500 km, with a total time duration of less than 1 ms. The first observation of an airglow enhancement was observed on board the Discovery Space Shuttle in 1992 \cite{boeck1992}. Since then, ELVES have been observed with various detectors, both ground and space-based \cite{sato2015} \cite{Newsome2010} \cite{chen2008}. 

This work will present recent results of the detection of ELVES by Mini-EUSO. In particular, the detector proved efficient in observing multiple ELVES, characterized by the occurrence of pairs of ELVES in rapid succession.

\section{The Mini-EUSO telescope}

\begin{figure}[!h]
\centering
\includegraphics[scale=0.2]{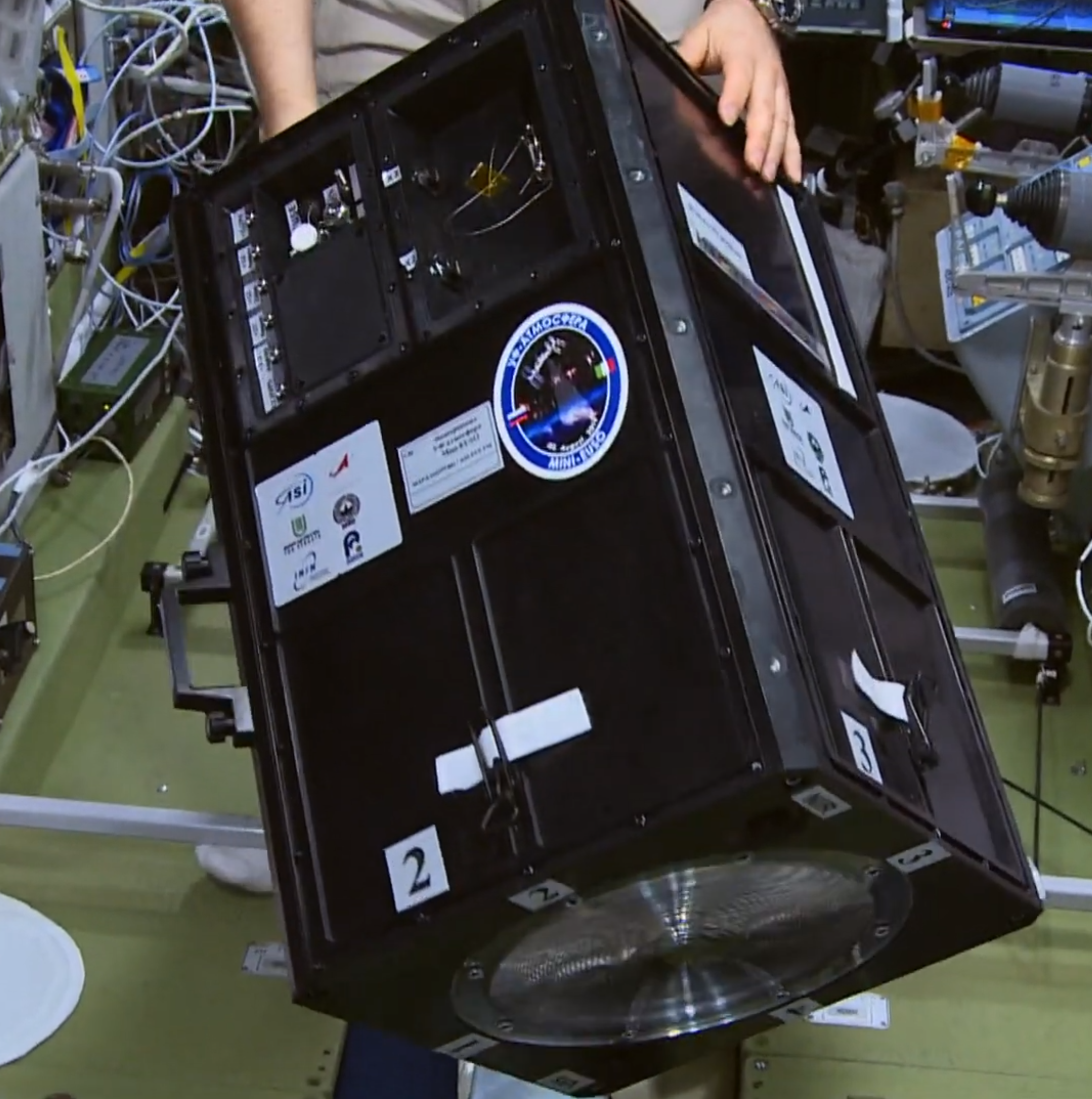}
\caption{The Mini-EUSO telescope on board the International Space Station, before being allocated on the nadir-pointing, UV transparent window of the Zvezda module.  The telescope is connected to the window via a mechanical adapter flange, here not shown.}
\label{fig:miniEUSO_ISS}
\end{figure}

The Mini-EUSO telescope ($37\times37\times62$~cm$^3$) observes the Earth through a UV-transparent window located in the Zvezda module of the International Space Station (Figure \ref{fig:miniEUSO_ISS}). 
The optical system of Mini-EUSO consists of two Poly(methyl methacrylate)(PMMA) Fresnel lenses, each measuring $25\;\text{cm}$ in diameter. These lenses improve light collection and have been specifically chosen for their suitability in space applications. The Focal Surface (FS) is composed of a matrix of $6\times6$ Multi-Anode Photomultiplier Tubes (MAPMTs, Hamamatsu Photonics R11265-M64).  Each MAPMT is a 64-channel ($8\times8$ pixels) device, providing a total of 2304 pixels. The active area of each MAPMT is $23\times23\;\text{mm}^2$, with individual pixel dimensions of approximately $2.9\times2.9\;\text{mm}^2$. The MAPMTs are grouped in Elementary Cells (ECs) of $2\times2$ MAPMTs. The spectral response of the telescope reaches a quantum efficiency of $35-45\%$ for light in the 290 to 430 nm wavelength range. A PMT fail-safe mechanism is implemented in Mini-EUSO, that reduces the gain (or turns off completely) PMTs exposed to bright light sources. The UV measurement is complemented by ancillary near-infrared and visible cameras. Overall, the telescope has $2.5\;\mu\text{s}$ temporal resolution and single photon detection capabilities. 

Mini-EUSO was launched on August 2019 and the first observations took place in October of the same year. The square field of view of the telescope is of $44^\circ$ and the spatial resolution of $6.3\times6.3\; \text{km}^2$ on ground level, this varying slightly with the altitude of the International Space Station and the pointing direction of each pixel. Being located in the middle of the Zvezda module, the detector is usually installed during onboard night-time. A data-taking session is usually about 12 hours long, during which the instrument automatically detects the day/night transitions. The
data gathered in a session usually range between 3 and 5 hours. 

\section{ELVES observed by Mini-EUSO}

\begin{figure} [!h]
    \centering
\includegraphics[width=\textwidth]{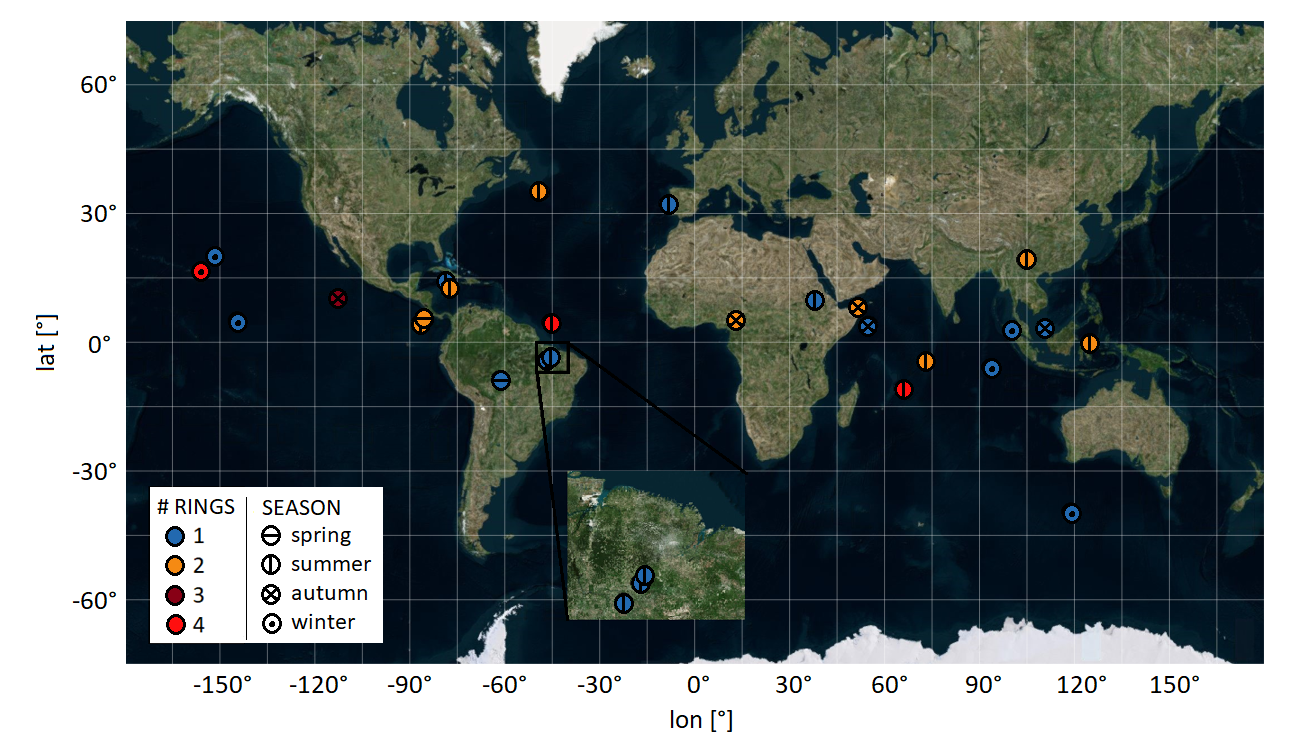}
    \caption{Location (latitude-longitude) of the ELVEs detected with Mini-EUSO. Most of them are distributed in the equatorial region. Three events in East Brazil occurred at an interval of 18 and 5 seconds (second from first and third from second respectively) one from the other.}
    \label{fig: Location}
\end{figure}

Mini-EUSO detected about 30 single and multiple-ringed ELVES so far. 
ELVES are traced by Mini-EUSO in their horizontally expanding, fast ring-shaped light emission. They appear as bright arcs of light on the telescope's focal surface, expanding over time. In general, several $2.5 \;\mu \text{s}$ time frames are associated with each event, having ELVES a duration of hundreds of $\mu \text{s}$. The typical total duration of the detected events reaches a maximum of about 200 frames, which means a total duration of about 0.5 ms, consistent with results reported in the literature. ELVES detected by Mini-EUSO are located mostly in the equatorial region (Figure \ref{fig: Location}), with about 2/3 of the observed events occurring over the ocean. This result is consistent with previous global occurrences of ELVES, including that by ISUAL in 2008, which reported how these events occurred mostly over the sea \cite{chen2008}. 

\begin{figure}
\centering
\includegraphics[width=1\textwidth]{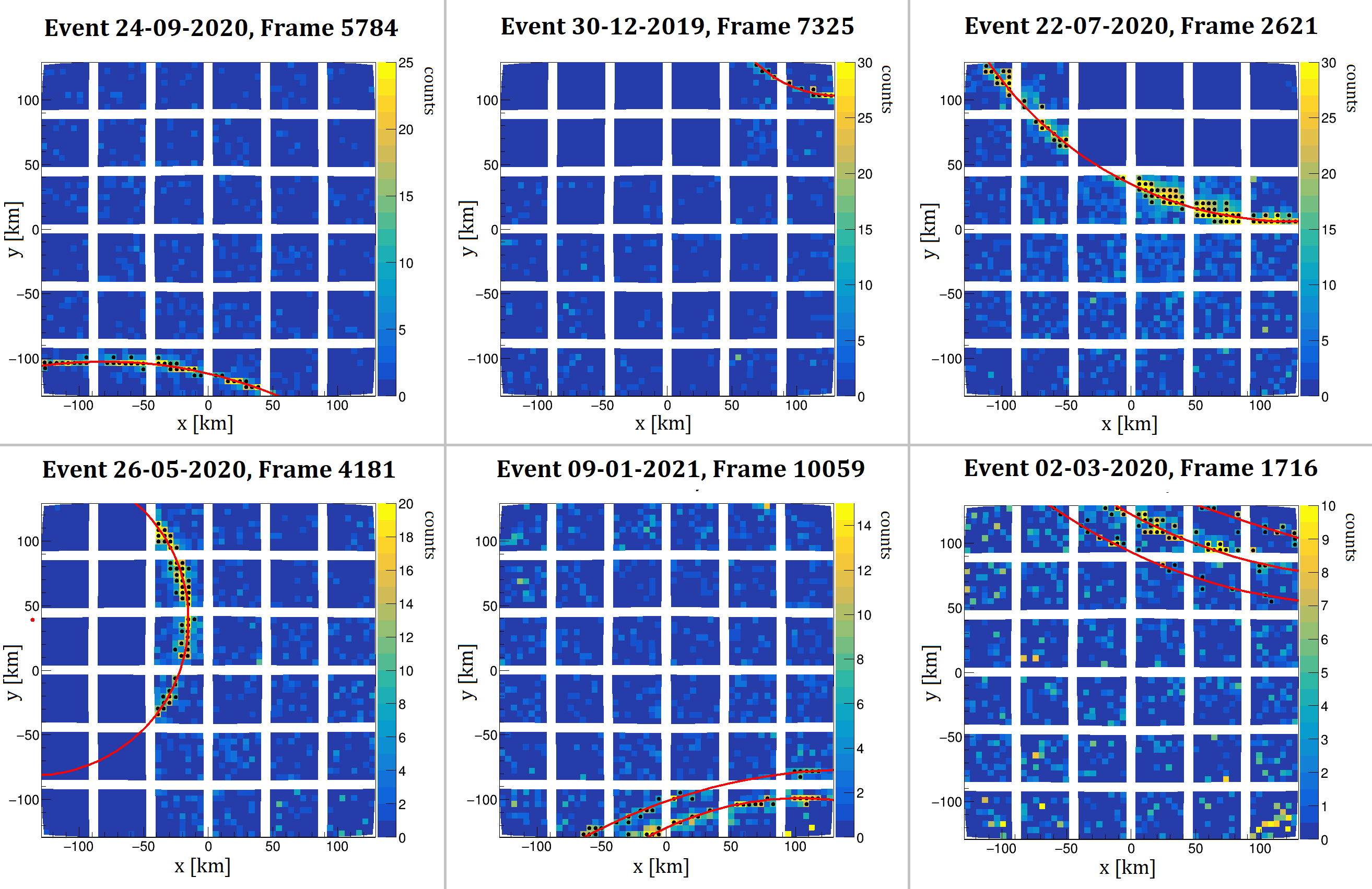}
\caption{One 2.5 $\mu s$ acquisition frame for a sample of ELVES observed by Mini-EUSO. ELVES appear as rings of light on the telescope's focal surface. From left to right, top to bottom: a, b, c) a single ring event; d) an ELVE with the center very close to the FoV of the focal surface; e) a double ring event, with an inter-ring distance of 50 $\mu s$; f) a multiple ring event with an inter-ring distance of 83 and 95 $\mu s$. The red lines represent the result of an iterative circle-fitting algorithm applied to each frame. The pixels selected by the algorithm and whose coordinates are used to fit the circle (red line) are marked with black dots. In all frames, color denotes counts/GTU (1 GTU = 2.5 $\mu$s). For events b, c, and d, some EC units are turned off due to the Mini-EUSO fail-safe mechanism.}
\label{fig-FrameFitCircle}        
\end{figure}

Using dedicated algorithms, the morphological characteristics of the ELVES detected by Mini-EUSO have been thoroughly investigated. The analysis carried out so far included modeling the arcs of light that are visible in the telescope images. Mini-EUSO has the advantage of observing the expanding rings of light from the nadir point of view, allowing a better geometrical reconstruction of the events. Assuming that ELVES propagate in the ionosphere as circles of increasing radius, a circle-fitting algorithm has been applied to the images of Mini-EUSO, to fit with a circle the visible arc of an ELVE ring on the focal surface and determine the position of the center of the ELVE in the ionosphere. The algorithm is iterative and starts from already existing solutions in literature \cite{Ladron2011}, adapted to the analysis of Mini-EUSO images. Some results can be appreciated in Figure \ref{fig-FrameFitCircle}. The developed procedure turns out to be fast and robust, being able to converge with outliers in the image and even with small (or incomplete) arc data. The algorithm works well also for multiple events, being able to recognize the presence of multiple ELVES in the same image by fitting each circle arc independently.

\begin{figure}
    \centering
    \includegraphics[width=\textwidth]{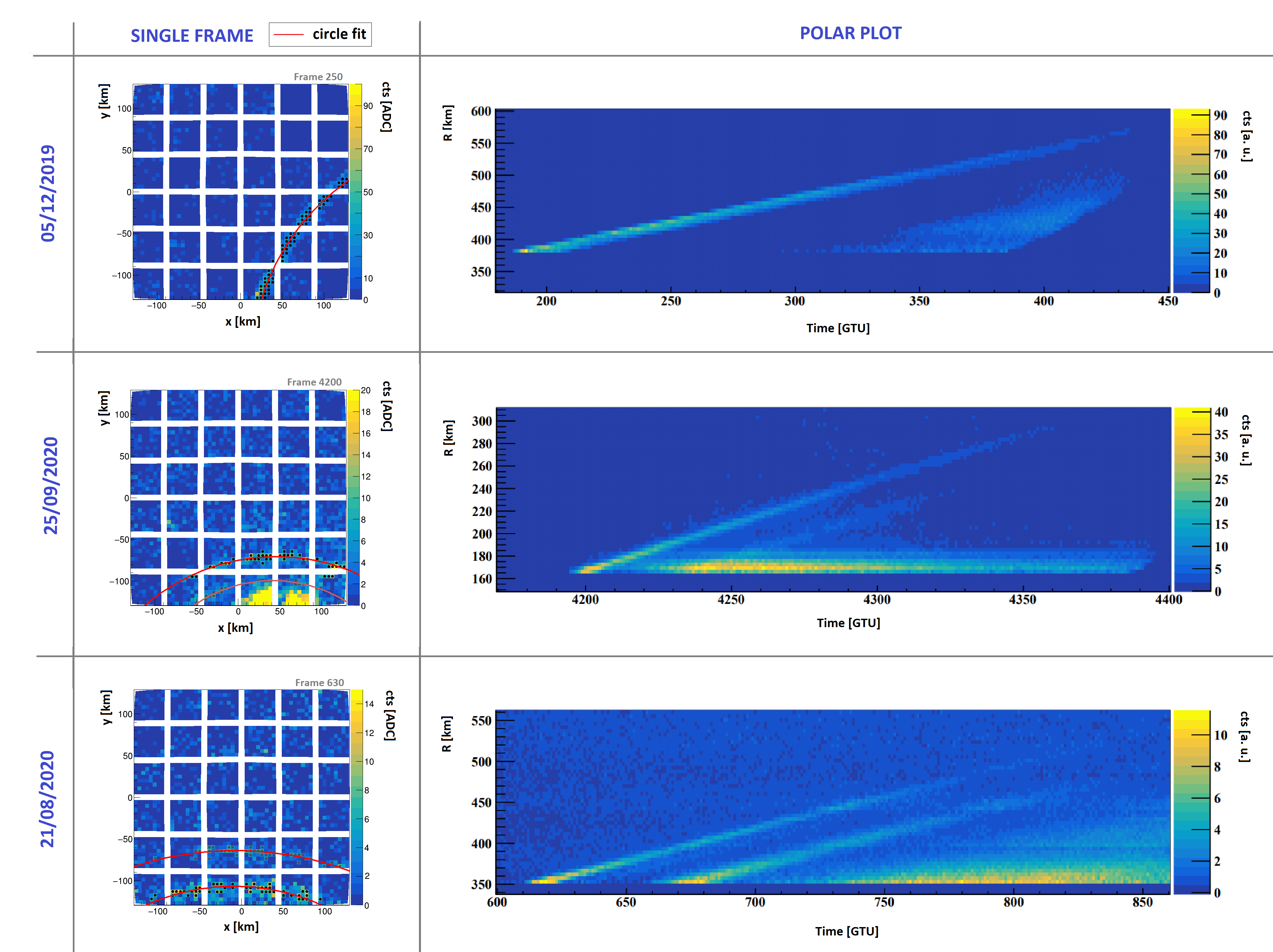}
    \caption{One 2.5 $\mu s$ acquisition frame (left column) and radius versus time distribution (right column) for a sample of ELVES observed by Mini-EUSO. The acquisition frames are represented following the same procedure described in Figure 3. In the R(t) distributions (or polar plots), ELVES appear as high-count inclined bands. From top to bottom: a) a single ring event; b) a double ring event, with a brightness concurrent with the ELVE, static in time; c) a double ringed event, with a brightness following the ELVE and expanding at the same velocity.}
    \label{fig: ICRC2}
\end{figure}

Once the position of the ELVE event in the ionosphere is computed, it is possible to calculate the total number of photoelectron counts detected at a given time t as a function of the radial distance R from the center.  Some distributions are shown in Figure \ref{fig: ICRC2}. With this representation, ELVES are visualized as high-count inclined bands. The velocity of propagation of the ring can be estimated by looking at the correspondent $R(t)$ distribution. Considering the propagation of the electromagnetic wave from the lightning to the ionosphere, the radius of the ELVE as a function of time $R(t)$ is of the form:

\begin{equation}
R(t)=\sqrt{c^2(t-t_0)^2-(h-h_0)^2}
\end{equation}

where $c$ is the speed of light; $h$ is the altitude of the ionosphere, and $h_0$ is the altitude of the lightning discharge that causes the event. For all the detected events, the velocity of propagation of the rings is consistent with light velocity. 

Finally, a spherical luminosity has been observed for some of the events detected by Mini-EUSO, both concurrent with and following the expansion of the ELVES (Figure \ref{fig: ICRC2}). The short temporal gap between the two phenomena suggests that the observed luminosity might be indicative of a halo. A more comprehensive analysis will be conducted in upcoming works. 

\section{Conclusions}

This work presents recent results on the observation of ELVES using the Mini-EUSO telescope. As expected, the majority of the events have been observed in the Earth's equatorial region and over the ocean. The temporal and spatial resolutions of the instrument allow for a detailed study of the morphological structure of ELVES, enhanced by the fact that the telescope observes the expanding rings of light from a nadir point of view. In this context, dedicated algorithms have been developed to reconstruct the circular structures of the expanding light rings and to estimate their propagation speed, which in all cases is consistent with the speed of light. Among the observed events, there are multiple events with two or more rings of light propagating simultaneously on the focal surface of the telescope. A more comprehensive analysis of their characteristics is currently underway. 

\section{Acknowledgments}

This work was supported by the Italian Space Agency through the agreement n. 2020-26-
Hh.0, by the French space agency CNES, and by the National Science Centre in Poland grants
2017/27/B/ST9/02162 and 2020/37/B/ST9/01821. This research has been supported by the Interdisciplinary
Scientific and Educational School of Moscow University “Fundamental and Applied
Space Research” and by Russian State Space Corporation Roscosmos. The article has been prepared
based on research materials collected in the space experiment “UV atmosphere”. We thank the
Altea-Lidal collaboration for providing the orbital data of the ISS.

\input{JEM-EUSO_Authors_July2023_final_v2.tex}

\end{document}

%% file: JEM-EUSO_Authors_July2023_final_v2.tex
\newpage
{\Large\bf Full Authors list: The JEM-EUSO Collaboration\\}
%{\scriptsize (author-list as of July 15th, 2023 with reorganized affiliations)} \hspace{0.6cm}
%{\scriptsize (version  \today{} \currenttime{})}
%\vspace*{0.5cm}

\begin{sloppypar}
{\small \noindent
S.~Abe$^{ff}$, 
J.H.~Adams Jr.$^{ld}$, 
D.~Allard$^{cb}$,
P.~Alldredge$^{ld}$,
R.~Aloisio$^{ep}$,
L.~Anchordoqui$^{le}$,
A.~Anzalone$^{ed,eh}$, 
E.~Arnone$^{ek,el}$,
M.~Bagheri$^{lh}$,
B.~Baret$^{cb}$
% ...
}
\end{sloppypar}
\vspace*{.3cm}

%%\newpage
{ \footnotesize
\noindent
% Algeria - 2 institutes
% ...
$^{cb}$ Universit\'e de Paris, CNRS, AstroParticule et Cosmologie, F-75013 Paris, France\\
% ...
$^{ed}$ Istituto Nazionale di Fisica Nucleare - Sezione di Catania, Italy\\
% ...
$^{eh}$ INAF - Istituto di Astrofisica Spaziale e Fisica Cosmica di Palermo, Italy\\
% ...
$^{ek}$ Istituto Nazionale di Fisica Nucleare - Sezione di Torino, Italy\\
$^{el}$ Dipartimento di Fisica, Universit\`a di Torino, Italy\\
% ...
$^{ep}$ Gran Sasso Science Institute, L'Aquila, Italy\\
% Japan - 7 institutes 
% ...
$^{ff}$ Nihon University Chiyoda, Tokyo, Japan\\ 
% ...
$^{ld}$ University of Alabama in Huntsville, Huntsville, AL, USA\\
$^{le}$ Lehman College, City University of New York (CUNY), NY, USA\\
% ...
$^{lh}$ Georgia Institute of Technology, USA\\
% ...
}